\documentclass[runningheads]{svmult}

\usepackage{makeidx}   % allows index generation
\usepackage{graphicx}  % standard LaTeX graphics tool
\usepackage{subeqnar}  % subnumbers individual equations
\usepackage{multicol}  % used for the two-column index
\usepackage{physprbb}  % modified textarea for proceedings,
\makeindex             % used for the subject index
\begin{document}
\title*{The cosmic saga of $^3$He}
\toctitle{The cosmic saga of $^3$He}
\titlerunning{The cosmic saga of $^3$He}
\author{Daniele Galli}
\authorrunning{D. Galli}
\institute{INAF-Osservatorio Astrofisico di Arcetri, Largo E. Fermi 5, 50125 Firenze, ITALY}

\maketitle              % typesets the title of the contribution

\begin{abstract}
We recall the emergence of the ``$^3$He problem'', its currently accepted
solution, and we summarize the presently available constraints on models
of stellar nucleosynthesis and studies of Galactic chemical evolution
from measurements of the abundance of $^3$He in the Galaxy.
\end{abstract}

\section{In the beginning was tralphium}

The isotope $^3$He probably first entered the astrophysical arena in 1949
with the (unpublished) calculations of Fermi \& Turkevich concerning the
chemical evolution of the first half-hour of the Universe.  The names
``tralphium'' and ``tralpha particles'' invented by George Gamow for this
isotope and its nuclei, have survived only in his humourous description
of nucleogenesis: {\it And God said: ``Let there be mass three.''
And there was mass three. And God saw tritium and tralphium, and they
were good''}. And so on to transuranium elements, with Fred Hoyle's help
to bridge the gap at mass five (Kragh~1996). The rough estimate of by
Fermi \& Turkevich ($^3{\rm He}\simeq 10^{-2}$ by mass) was later refined
by more detailed calculations, like e.g. those of Wagoner, Fowler, \&
Hoyle~(1967) who showed that $^3$He could be produced at levels comparable
to its terrestrial abundance ($\sim 5\times 10^{-5}$ by mass) during
the evolution of a ``universal fireball or a supermassive object'',
or, as we say today, in the big bang .  Thus, at least in principle,
the abundance of $^3$He could be used (together with D, $^4$He and Li)
to test theoretical predictions, and, in particular, to constrain the
baryon density of the Universe. Having gained the special status of
``cosmic baryometer'' and caught the attention of cosmologists, the
interest in $^3$He spread rapidly in the astronomical community.

\section{Trouble ahead}

At around the same time, Iben (1967) and Truran \& Cameron (1971) showed
that ordinary stars produce $^3$He in the ashes of hydrogen burning
by $p$--$p$ cycle on the main sequence. They found that the stellar
production of $^3$He roughly scales as $M^{-2}$, where $M$ is the mass
of the star, indicating that low-mass stars ($M\simeq 1$--3 $M_\odot$)
are the dominant site of $^3$He production in the Galaxy.  Problems
followed soon, when Rood, Steigman, \& Tinsley (1976) incorporated
the stellar production of $^3$He in simple models of Galactic chemical
evolution, and found the predicted present-day abundances to be larger
by orders of magnitude than the value measured in samples of gas-rich
meteorites, representative of interstellar medium abundances at the
time of formation of the Sun. The paper by Rood, Steigman, \& Tinsley
(1976) marked the first appearance of the ``$^3$He problem''. However,
additional observations of $^3$He in the Galaxy were needed to confirm
the exent of the discrepancy.

\section{Observing $^3$He}

%In the last two decades, a considerable collection of $^3$He abundance
%determinations has been assembled with a variety of methods in different
%Galactic environments, H{\sc ii} regions, planetary atmospheres, neutral
%and ionized local interstellar medium, and planetary nebulae. This body
%of data constitutes the empirical foundation for the existence of the
%``$^3$He problem'', and provides strong constraints for any possible
%solution. It is therefore important, and hopefully useful for Galactic
%evolution modellers, to summarize here the main results.

Radioastronomers first learned of $^3$He in 1955 at the fourth I.A.U.
Symposium in Jodrell Bank, when the frequency of the hyperfine $^3$He$^+$
line at 8.666 GHz (3.46 cm) was included by Charles Townes in a list of
``radio-frequency lines of interest to astronomy'' (Townes~1957). The
line was (probably) detected for the first time only twenty years
later, by Rood, Wilson \& Steigman~(1979) in W51, opening the way to
the determination of the $^3$He abundance in the interstellar gas of our
Galaxy via direct (although technically challenging) radioastronomical
observations. In the last two decades, a considerable collection of
$^3$He$^+$ abundance determinations has been assembled in H{\sc ii} regions
and planetary nebulae. The relevance of these results will be discussed in 
Sect.~4 and 5 respectively.

For many years, meteorites have provided the only means to determine
the abundance of $^3$He in protosolar material.  The values obtained by
mass spectroscopy techniques in the so-called ``planetary'' component
of gas-rich meteorites have been critically examined by Geiss~(1993)
and Galli et al. (1995). The latter recommend the value 
$^3$He/$^4$He$=(1.5\pm
0.1)\times 10^{-4}$.  
%Exposure to the $^3$He rich solar wind, isotopic
%fractionation when He was incorporated in the bodies from which the
%meteorites formed, and other processes, may cast some doubt on the
%relevance of the meteoritic value as an indication of the interstellar
%abundance at the time of formation of the Sun.  Fortunately, the
The meteoritic value has been confirmed by {\it in situ} measurement of
the He isotopic ratio in the atmosphere of Jupiter by the Galileo
Probe Mass Spectrometer. The isotopic ratio obtained in this way,
$^3$He/$^4$He$=(1.66\pm 0.04)\times 10^{-4}$ (Mahaffy et al. 1998),
is slightly larger than, but consistent with, the ratio measured in
meteorites, reflecting possible fractionation in the protosolar gas in
favor of the the heavier isotope, or differential depletion in Jupiter's
atmosphere.

The He isotopic ratio in the present day local ISM (inside and
beyond the heliosphere at 3--5 AU from the Sun) has been determined
by two recent space experiments, and the two results agree within
the uncertainties. Helium atoms entering the solar system from the
surrounding interstellar cloud and ionized deep inside the heliosphere
(the so-called ``pick-up'' ions), analyzed by the Solar Wind Ion
Composition Spectrometer on the Ulysses spacecraft, show an isotopic ratio
$^3$He/$^4$He$=(2.5^{+0.7}_{-0.6})\times 10^{-4}$, with the uncertainty
resulting almost entirely from statistical error (Gloecker \& Geiss
1998). In the Collisa experiment on the russian space station Mir, on
the other hand, samples of the local {\it neutral} ISM were collected
on thin metal foils exposed to the flux of interstellar particles,
and later analyzed in terrestrial laboratories. The He isotopic ratio
measured in this way is $^3$He/$^4$He $=(1.7 \pm 0.8) \times 10^{-4}$
(Salerno et al. 2003).

\section{The age of reason}

The old problems have now largely been overcome, and new ones
have appeared.  As for the cosmological implications, thanks to the
continuing effort of Rood and collaborators over more than two decades
to determine $^3$He abundances in H{\sc ii} regions (see contribution
by Bania et al. in these proceedings), the usefulness of this isotope
as a cosmic baryometer has now been fully established. The trend (or
better, the absence of a trend) of $^3$He vs. metallicity in a sample
of about 40 H{\sc ii} regions reveals the existence of a ``$^3$He
plateau'' at $^3$He/H$=(1.1\pm 0.2)\times 10^{-5}$, similar in many
ways to the celebrated ``Li plateau''. The resulting baryon-to-photon
ratio $\eta_{10}=5.4^{+2.2}_{-1.2}$ (Bania, Rood \& Balser 2002) is
in agreement with other independent determinations of this fundamental
cosmological parameter.  After fifty years, the program started by Fermi
\& Turkevich's theoretical prediction of ``tralphium'' production in
the early universe seems to have reached its fulfillment.

%early attempts to find elegant nuclear physics solutions (Galli
%et al.~1994), were found to be very unlikely in recent laboratory
%experiments (Bonetti et al.~1999).

As for the discrepancy between observed abundances of $^3$He and
the predictions of models of Galactic chemical evolution, the
natural explanation of the problem was found by Charbonnel~(1995)
and Hogan~(1995) in the existence of a non-standard mixing mechanism
acting in low-mass stars during the red-giant branch evolution or later,
leading to a substancial (or complete) destruction of all their freshly
produced $^3$He.  In this way, the ``$^3$He problem'' was reduced to
``just another'' isotopic anomaly, similar to those commonly observed
in the atmospheres of giant stars for elements like carbon and oxygen,
as originally suggested by Rood, Bania \& Wilson~(1984) almost ten
years earlier.  The characteristics of this mixing mechanism, and the
attempts to identify a physical mechanism responsible for its occurrence
(rotation?) have been nicely reviewed by Charbonnel~(1998), and will
not be repeated here.  For an impressive demonstration of the effects
of extra-mixing on the carbon isotopic ratio in globular cluster stars,
the reader should look at Fig.~2 of Shetrone~(2003).

\begin{figure}[t]
\begin{center}
\includegraphics[width=0.8\textwidth]{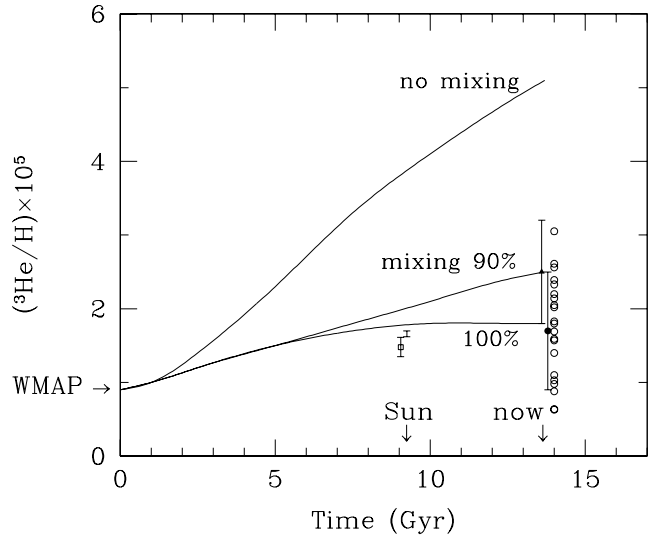}
\end{center}
\caption[]{Evolution of $^3$He/H in the solar neighborhood, computed
without extra-mixing ({\it upper curve}\/) and with extra-mixing in 90\%
or 100\% of stars $M<2.5~M_\odot$ ({\it lower curves}\/). The two arrows
indicate the present epoch (assuming a Galactic age of $13.7$~Gyr) and
the time of formation of the solar system 4.55~Gyr ago.  Symbols and
errorbars show the $^3$He/H value measured in: meteorites ({\it empty
squares}); Jupiter's atmosphere ({\it errorbar}); the local ionized ISM
({\it filled triangle}); the local neutral ISM ({\it filled circle});
the sample of ``simple'' H{\sc ii} regions ({\it empty circles}). Data
points have been slightly displaced for clarity. The He
isotopic ratios has been converted into abundances relative to hydrogen assuming a universal
ratio He/H$=0.1$.  See text for references.}
\label{evolution} 
\end{figure}

Fig.~\ref{evolution} (adapted from Fig.~4 of Romano et al.~2003) shows
the evolution of $^3$He/H in the solar neighborhood, computed with the
model of Tosi (2000) assuming the standard (without extra-mixing)
stellar yields of Dearborn, Steigman, \& Tosi (1996) and the
extra-mixing yields of Boothroyd \& Sackmann (1999) for 90\% ans 10\%
of stars with $M<2.5~M_\odot$ (see Galli et al.~1997 and Romano et
al.~2003).  Symbols and errorbars show the $^3$He/H value measured in:
meteorites (Galli et al.~1995); Jupiter's atmosphere (Mahaffy et
al.~1998); the local ionized ISM (Gloecker \& Geiss~1998); the local
neutral ISM (Salerno et al.~2003); the sample of ``simple'' H{\sc ii}
regions (Balser et al.~2002).  The primordial abundance of $^3$He
corresponding to the baryon-to-photon ratio determined by WMAP (Spergel
et al.~2003) is indicated by an attow at $t=0$.  Taken together, the
observational data support the hypothesis that negligible changes of
the abundance of $^3$He have occurred in the Galaxy during the past
4.55 Gyr. The failure of the standard $^3$He yields to account for the
measured abundances is not a peculiarity of the particular Galactic
model shown in Fig.~1, as the interested reader may see in Fig.~6 of
Tosi~(1998). It should be noted however that the discrepancy with the
observational data is rather model dependent.

It is evident from Fig.~\ref{evolution} that consistency with the observed
abundance of $^3$He in the Galaxy is achieved only if the fraction of
low-mass stars ($M < 2.5~M_\odot$) undergoing extra-mixing is larger than
$\sim 90$\%, assuming the $^3$He yields of Boothroyd \& Sackmann (1999).
Thus, to solve the $^3$He problem in terms of extra-mixing in low-mass
stars, the vast majority of them (90\%--100\%) must be affected by this
phenomenon (Galli et al.~1997).  The same conclusion has been reached
independently by Charbonnel \& do Nascimento~(1998) on the basis of the
statistics of carbon isotopic ratios in a sample of red-giant stars with
accurate Hipparcos parallaxes.

\begin{figure}[t]
\begin{center}
\includegraphics[width=0.8\textwidth]{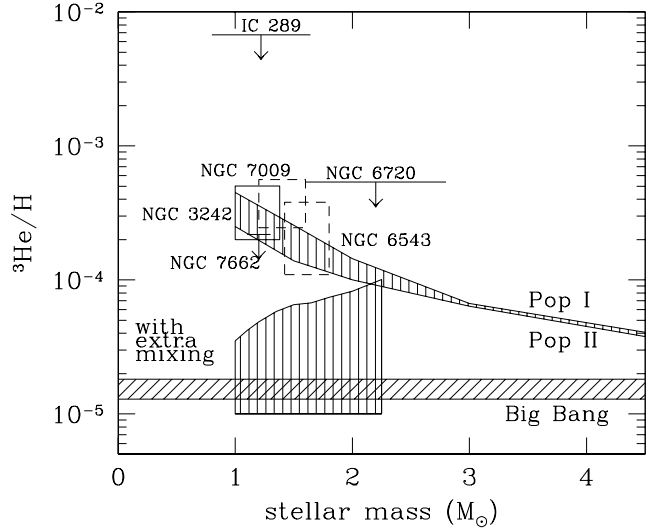}
\end{center}
\caption[]{Abundance of $^3$He vs. main-sequence masses (determined by
Galli et al.~1997) for the six PN of the sample of Balser et al.~(1997)
and Balser, Rood, \& Bania~(1999).
The curves labeled ``Pop~I'' and ``Pop II'' show the ``standard'' abundance
of $^3$He computed by Weiss, Wagenhuber, \& Denissenkov~(1996) for two
metallicities. The curves
labeled ``with extra-mixing'' show the results of stellar
nucleosynthesis calculations with deep mixing by Boothroyd \&
Sackmann~(1999) (upper curve) and the equilibrium value
$^3{\rm He/H}=10^{-5}$ for $M<2.5~M_\odot$ (lower curve).}
\label{pn}
\end{figure}

\section{A final touch: planetary nebulae}

The most direct, model independent, way to test the validity of the
mixing solution is to measure the $^3$He abundance in the ejecta of
low-mass stars, i.e. in planetary nebulae (PNe). The search for $^3$He
in the ejecta of PNe via the 8.667 GHz spin-flip transition of
$^3$He$^+$, painstakingly carried out by Rood and coworkers at the
Green Bank radiotelescope since 1992 (see summary of results in Balser
et al.~1997), has produced so far {\em one solid detection} (NGC~3242,
see Rood, Bania, \& Wilson~1992; confirmed with the Effelsberg
radiotelescope by Balser, Rood, \& Bania~1999), {\em two tentative
detections} (IC~289, NGC~6720) and {\em three upper limits} (NGC~7662,
NGC~6543, NGC~7009). One more detection has been recently obtained with
the NRAO VLA in the PN J320 (Balser et al., these proceedings).  All
these objects can be placed in the progenitor mass--$^3$He diagram (see
details of the procedure in Galli et al.~1997), and compared with the
predictions of stellar models with and without extra-mixing
(Fig.~\ref{pn}). Ironically enough, the position of the six PNe
definitely supports the standard $^3$He yields, in particular the
(only) solid detection of the sample, NGC~3242. Although the
statistical significance of the sample is questionable, and selection
biases are certainly present, the only way to reconcile Fig.~1 with
Fig.~2 is to conclude that most, if not all, the PNe shown in
Fig.~\ref{pn} belong to the $10$\% (or less) of low-mass stars which
did {\em not} experience extra-mixing. 

\section{Conclusions}

We have learned many things about Gamow's tralphium since 1949. A personal
selection includes: (1) the abundance of $^3$He has not changed significantly
over $\sim 14$~Gyr of Galactic evolution, which is remarkable; (2) it has
not changed not because nothing happened, but because two independent
processes of opposite sign and equal magnitude were at work, which is
{\it truly} remarkable; (3) one object does not make a statistically
significant sample; (4) many objects do not make it either, if they are
selected carefully enough.

%(5) physical problems often have elegant solutions but Nature not 
%always has the good taste of choosing them.

\bigskip

\noindent {\bf Acknowledgements.} This work is supported by a grant COFIN~$2002027319~003$.
It is a pleasure to thank the members of
the international ``$^3$He community'' for sharing their results and
enthusiasm on our beloved isotope. Special thanks to Monica Tosi for
providing an enlightening guidance to the intricacies of $^3$He, and an
exquisite company at many conferences on the light elements.

\end{document}